\documentclass[%
 aps,reprint,
 prl,
 amsmath,amssymb,
]{revtex4-2}

\usepackage{graphicx}
\usepackage{dcolumn}
\usepackage{bm}
\usepackage{xcolor}

\newcommand{\commentColor}{\textcolor{black}}

\begin{document}


\title{Enhanced performance in quasi-isodynamic max-$J$ stellarators with a turbulent particle pinch}

\author{G. G. Plunk, A. G. Goodman, P. Xanthopoulos, P. Costello, H. M. Smith, K. Aleynikova, C. D. Beidler, M. Drevlak, S. Stroteich and P. Helander}
\affiliation{Max-Planck-Institut f{\"u}r Plasmaphysik, 17491 Greifswald, Germany}



\date{\today}

\begin{abstract}
 Recent stellarator reactor designs demonstrate mostly outward turbulent particle transport, which, without advanced fueling technology, inhibits the formation of density gradients needed for confinement.  We introduce ``SQuID-$\tau$'', a self-fueling quasi-isodynamic stellarator capable of sustaining density peaking through inward particle transport caused by turbulence. Temperature and density profile predictions based on high-fidelity gyrokinetic simulations demonstrate enhanced performance, significantly relaxing constraints on the size and magnetic field strength for reactor designs.
\end{abstract}

\maketitle

The neoclassically optimized stellarator experiment Wendelstein 7-X (W7-X) \cite{Klinger2019, Nature21} achieves its best energy confinement once a peaked density profile is established within the plasma volume \cite{Bozhenkov2020}.  Such enhanced confinement is theoretically expected in a quasi-isodynamic (QI) stellarator that possesses the max-$J$ property \cite{Helander2013, Alcuson2013, rodriguez2024maximum}, and it is therefore sensible to design stellarator fusion reactors based on this principle.  

However, although peaked density profiles in W7-X are sustained through an inward turbulent particle transport (``pinch'') of the main plasma species \citep{thienpondtPreventionCoreParticle2023, ford2024turbulence, bannmann2024particle}, recently proposed reactor-relevant QI stellarator designs \citep{Sánchez_2023, goodmanQuasiIsodynamicStellaratorsLow2024b, Lion2025stellaris, Hegna_2025} exhibit, at most, only a weak pinch \citep{Garcia-Regana_TurbulenceCIEMATQI_2025, Guttenfelder_Infinity2_2025}. In order to achieve good confinement, the operation of reactors based on such designs would rely heavily on advanced particle fueling techniques, like cryogenic pellets or neutral beams, that are unfortunately associated with neoclassically-driven \citep{romba2025suppression} impurity accumulation that can pose a severe threat to performance \citep{warmer2016system}.  With impurity accumulation and particle fueling insufficiently understood, especially for core fueling with pellets in high-temperature reactor conditions \citep{pegourie2007pellet}, there is inherent risk in building such reactor designs.

These considerations motivate the search for well-optimized QI stellarator configurations that could achieve intrinsically enhanced confinement via a sufficiently strong turbulent particle pinch.  This ``self-fueling'' could be combined with a certain degree of external fueling to obtain even greater performance than possible with the individual effects acting alone, and mastering the  balance could be essential to find a path toward steady-state high-performance reactor operation.

\begin{figure}
    \centering
    \includegraphics[width=0.45\textwidth]{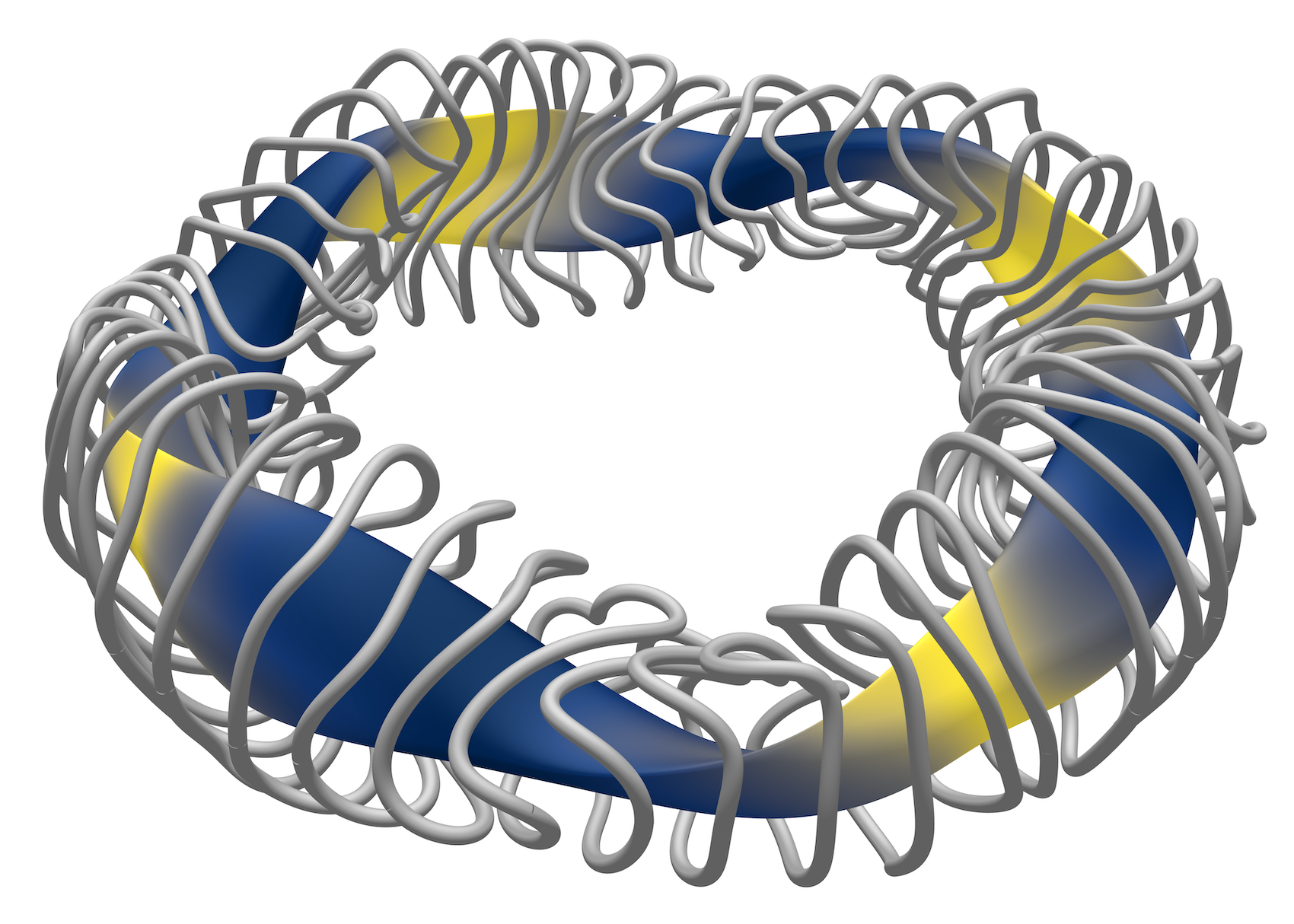}
    \caption{The self-fueling quasi-isodynamic stellarator design SQuID-$\tau$, shown with its filamentary coil set.}
    \label{fig:tau}
\end{figure}

Extending the methodology behind the SQuID line of stellarator designs \cite{goodmanQuasiIsodynamicStellaratorsLow2024b}, we present the first max-$J$ QI configuration with a strong turbulent particle pinch.  The new stellarator design, which we call ``SQuID-$\tau$" (see Fig. \ref{fig:tau}), also fulfills other key properties, even at modest reactor scales (volume of 1450 m$^3$): zero collisionless fast-particle losses at $\rho=0.5$ for volume-average $\langle\beta\rangle\geq 2\%$, Mercier stability at $\langle\beta\rangle \lesssim 7\%$ throughout the plasma in fixed-boundary, bootstrap current $\lesssim10\textrm{ kA}$ at $\langle\beta\rangle \sim 2\%$ (density and temperatures consistent with $3$ GW of fusion power), and better coil compatibility \cite{Kappel_2024} than previous QI reactor candidates \cite{goodmanQuasiIsodynamicStellaratorsLow2024b,Lion2025stellaris}.  The coil set depicted reproduces the field with a relative error of less than $0.5\%$ on average.\footnote{\commentColor{All calculations in this paper were performed with the fixed-boundary equilibria, neglecting any corrections due to coils.}} Combining high-fidelity gyrokinetic turbulence simulations with transport solvers to predict plasma profiles, we show enhanced confinement of SQuID-$\tau$ compared to that of a previous SQuID design, the configuration underlying the Stellaris reactor concept \citep{Lion2025stellaris}.  We also demonstrate that enhanced confinement, acting in concert with density peaking, has very positive implications for a potential fusion reactor, in terms of the necessary levels of magnetic field strength and plasma volume to achieve a desired fusion gain.

{\bf Particle pinch in max-$J$ devices.}  Quasi-linear theory predicts a tendency for max-$J$ stellarators to have an outward flux of particles \citep{Helander_Zocco_2018}.  However, W7-X, an approximately max-$J$ stellarator, exhibits a turbulent particle pinch \citep{thienpondtPreventionCoreParticle2023, ford2024turbulence, bannmann2024particle}, as evidenced by the shape of observed density profiles, which are not hollow as they would be under the influence of neoclassical transport \citep{Beidler_2018}.

To see how such a pinch might arise, we can examine the quasi-linear particle flux $\Gamma_e^\mathrm{GK} = -k_\alpha \operatorname{Im} \left\langle  \phi^*\int g_e \mathrm{d}^3 v \right\rangle$, where $g_e$ is the linear, non-adiabatic response of the electrons, $\phi^*$ is the (conjugate) electrostatic potential, and $\langle \ldots \rangle$ denotes a flux-surface average. 
Due to the intrinsic ambipolarity of gyrokinetic particle transport, we need only consider the response of the electron species, and we decompose the particle flux $\Gamma_e^\mathrm{GK} = \Gamma_e^{tr} + \Gamma^p_e$ into trapped and passing electron contributions, respectively.

In the limit $\omega/(v_{Te} k_\|) \ll 1$, where the mode frequency is much smaller than the electron transit frequency, the dominant contribution to the trapped electron response is that of \cite{helanderQuasilinearParticleTransport2018a}. Taking the `non-resonant' limit of $\omega_{de}/\omega \sim \omega/\omega_{*e} \ll 1$, where $\omega_{de}$ is the electron drift frequency and $\omega_{*e}$ is the diamagnetic frequency, yields,
\begin{widetext}
\begin{equation}
    \Gamma_{e}^{tr} \frac{\mathrm{d}\ln n}{\mathrm{d}\psi} = - \frac{e^2 n}{2 \mathcal{V} T_e^2 }\int_{1/B_{\mathrm{max}}}^{1/B_{\mathrm{min}}} \mathrm{d}\lambda \sum_j \tau_{B,j} |\bar \phi_j|^2
    \frac{\omega_{*e}^2 \gamma}{(\omega_r^2 + \gamma^2)^2}\left[ 3 \tilde{\omega}_{de}(\lambda)\omega_r (1 + \eta_e) + (\omega_r^2 + \gamma^2)\right],
\end{equation}
\end{widetext}
where $\tilde{\omega}_{de}(\lambda) = \overline{ \hat \omega_{de}(l)(1 - \lambda B/2)}$, $\hat \omega_{de}(l)$ captures the geometric dependence of the electron curvature drift, $\lambda = v_\perp^2/(v^2B)$, $\overline{(\ldots)}_j$ denotes an average over the trapped particle orbit in a trapping-well indexed by $j$, with a bounce time $\tau_{B,j}$, and $\mathcal{V} = \int d l/B$. The plasma gradients enter explicitly via $\omega_{*e} = (-{k_\alpha T_e}/{e})({\mathrm{d}\ln n}/{\mathrm{d}\psi})$, $\eta_e = \mathrm{d} \ln T_e/ \mathrm{d}\ln n$. In a max-$J$ device, $\omega_{*e} \tilde\omega_{de} <0$, such that the trapped electrons precess in the ion-diamagnetic direction. As a result, in the presence of modes driven by the ion temperature gradient (ITG), which also tend to propagate in the ion-diamagnetic direction, we have $\tilde{\omega}_{de}\omega_r >0$. This results in $\Gamma_e^{tr} \mathrm{d} \ln n/\mathrm{d} \psi <0$, such that the particle flux is against the density gradient, i.e., outwards.

To see how inward particle transport might nevertheless occur in a well-optimized max-$J$ stellarator, we note that, although the passing electron response is an order $\omega/(v_{Te} k_\|)$ smaller than the trapped electron response, its contribution to particle transport can compete with the trapped response if the population of trapped particles at the ITG mode location $\ell_0$ is sufficiently small $f_\mathrm{tr} (\ell_0) \sim \omega/(v_{Te} k_\|)$. This situation is actually to be expected in max-$J$ stellarators, as the minima of the magnetic field tend to be out of alignment with the areas of maximum ITG drive (``bad curvature'') \citep{prollResilienceQuasiIsodynamicStellarators2012, Helander2013}.  In this case, the passing contribution $\Gamma_e^{p}$ must be included at the same order as the trapped contribution,
\begin{widetext}
\begin{equation}
     \Gamma_e^{p} \frac{\mathrm{d}\ln n}{\mathrm{d}\psi} = - \frac{e^2 n  \omega_{*e}^2 }{2T_e^2 \sqrt{\pi} v_{Te} \mathcal{V}}\left(1 - \frac{\eta_e}{2} - \frac{\omega_r}{\omega_{*e}} \right)\int_0^{1/B_\mathrm{max}}\mathrm{d}\lambda\, \bigg|\int_{-\infty}^{\infty} \frac{\phi \,\mathrm{d}l}{\sqrt{1 - \lambda B}} \bigg|^2. \label{eq:passing-particle-flux}
\end{equation}
\end{widetext}
Indeed, we see that, for $\eta_e > 2(1 - \omega_r/\omega_{*e})$, the passing electrons provide an inward contribution to the particle flux \citep{antonsenInwardParticleTransport1979a, hallatschek2005giant, jenko2005_heat_and_particles}. To optimize a max-$J$ configuration to exhibit a particle pinch for ITG-dominated turbulence, $\Gamma_e^{tr}$ must be made small compared to $\Gamma_e^p$.  
It is this insight that guided the optimization of the SQuID-$\tau$ design, while controlling for other known influences on the strength of turbulence (such as flux compression, the max-$J$ property, and the turbulent zonal flow response) studied in previous publications \citep[see {\em e.g.}][]{SquidX2013}.  In what follows, we will demonstrate the turbulent particle pinch of SQuID-$\tau$, and overall implications for plasma performance.

{\bf Transport modeling.} The steady state temperature profile (with good thermal coupling $T_i = T_e = T$) can be obtained by solving a single heat transport equation

\begin{equation}
    \frac{1}{V'}\frac{\textrm{d}}{\textrm{d}\rho}\left( V'\sum_\sigma \langle (Q_\sigma^\mathrm{GK} + Q_\sigma^\mathrm{NC})\cdot\boldsymbol{\nabla}\rho \rangle \right) = P_\alpha + P_\mathrm{ext},\label{eq:heat-transport}
\end{equation}
where $P_\mathrm{ext}$ is the external heating power, $P_\alpha$ is the effective heating power generated by fusion-born alpha particles, $V' = dV/d\rho$ and $Q_\sigma^\mathrm{GK}$ and $Q_\sigma^\mathrm{NC}$ are the turbulent and neoclassical heat fluxes for species $\sigma = i,e$.  The density profile, established in a region of the plasma without particle sources, is obtained by simultaneously solving the steady state condition
\begin{equation}
\frac{d\ln T}{d\ln n} = \eta_\mathrm{crit}(\rho),\label{eq:eta-eqn}
\end{equation}
where the function $\eta_\mathrm{crit}(\rho)$ is obtained by setting the density gradient $a/L_n = -d\ln n/d\rho$, at each radial grid point, for a fixed value of $a/L_T = -d\ln T/d\rho$, such that the gyrokinetic particle flux $\Gamma^\mathrm{GK}$ is zero to within statistical uncertainty.  The use of a critical $\eta$ model is motivated by the criterion for passing electrons to contribute a pinch, stated after Eqn.~\ref{eq:passing-particle-flux}.

Using the GX \cite{Mandell24} code, we perform electrostatic turbulence simulations, assuming equal temperatures and gradients for ions and electrons, at five different radial locations, for different values of $a/L_T$, in order to construct a transport model for $\hat{Q}_\mathrm{tot} =  q_\mathrm{tot}^\mathrm{GK}/Q_{\mathrm{GB, i}}$, where $q_\mathrm{tot}^\mathrm{GK} = \sum_\sigma \langle Q_\sigma^\mathrm{GK}\cdot\boldsymbol{\nabla} r \rangle$, $r = a \rho$ and $Q_\mathrm{GB,i} = \rho_i^2 c_s n T/a^2$, with $\rho_i$ the ion Larmor radius, $c_s$ the sound speed and $a$ the minor radius.  The transport equations are then solved for $T$ and $n$ using this model.  This approach has the advantage of allowing efficient exploration of different operation scenarios \citep{Warmer_2018, Beurskens_2021}, as compared with coupling a gyrokinetic solver directly to a transport solver for each scenario separately \citep{qian-2022-t3d, navarro-2023-gene-tango}.

\begin{figure}
    \centering
    \includegraphics[width=0.45\textwidth]{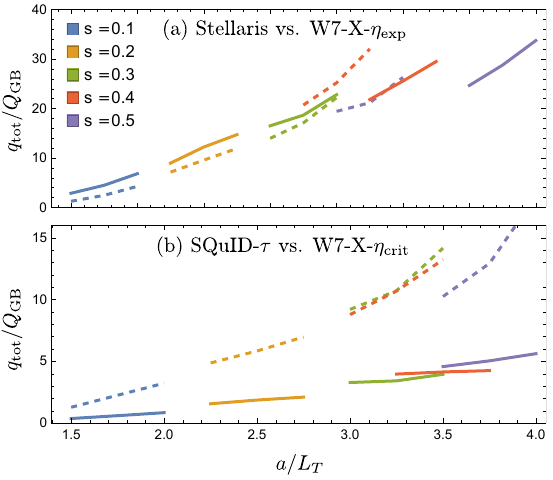}
    \caption{Total heat fluxes from GX simulations for Stellaris (top; dashed) and SQuID-$\tau$ (bottom; dashed), compared with those of W7-X standard configuration (solid) for $\eta(\rho) = \eta_\mathrm{exp}$ and $\eta(\rho) = \eta_\mathrm{crit}$, respectively.  Note that data appears in order of increasing $s = \sqrt{\psi/\psi_\mathrm{edge}}$, from left to right in the figures.}
    \label{fig:heat-fluxes}
\end{figure}

The results of the gyrokinetic simulations are shown in Figure \ref{fig:heat-fluxes}.  Values of $\eta_\mathrm{crit}$ for the Stellaris configuration (not plotted) are found to be in the range of $4$ to $5$ across the plasma radius, similar to observed values of $\eta$ in pure electron-cyclotron heated discharges of W7-X.  For this reason, we compare the heat fluxes of Stellaris against those of W7-X assuming experimentally motivated values of $\eta = \eta_\mathrm{exp}(\rho)$.  For SQuID-$\tau$, however, values of $\eta_\mathrm{crit}$ are in the range of $2.5-3$, which is closer to values of $\eta$ found in W7-X under higher performance scenarios, and we therefore compare its transport against W7-X simulations using $\eta = \eta_\mathrm{crit} \sim 2$.  This lower value of $\eta_\mathrm{crit}$, and therefore higher $a/L_n$, helps to explain the lower total transport of W7-X in this comparison.

It is the size of $\eta_\mathrm{crit}$ that we use to characterize the strength of the turbulent pinch, {\em i.e.} the lower its value, the stronger the pinch.  Figure \ref{fig:heat-fluxes} shows a significant reduction of the heat transport (sum of electron and ion thermal fluxes) in SQuID-$\tau$ as compared to Stellaris, for equal temperature gradient $a/L_T$, which relies on the peaked density gradients in SQuID-$\tau$, showing the benefit of the particle pinch optimization of SQuID-$\tau$.

{\bf Reference experimental scenarios.}  To obtain a global estimate of confinement and performance, we solve the transport equations in different operation scenarios.  We note, however, that first principles prediction of plasma profiles can be strongly affected by the choice of boundary conditions for the temperature at the transition between the core region (where Gyro-Bohm scaling prevails) and the plasma edge region.  As profile shapes and confinement scaling depends on the operating scenario, even within a single machine, it is important that the choice of boundary condition used for predicting the performance of a new stellarator design should be tied as closely as possible to the most relevant experimental data.  Accounting for the strong differences in the performance of the two SQuIDs (Fig.~\ref{fig:heat-fluxes}), we chose two different reference experimental W7-X discharges for comparison.  To assess the confinement quality, we use the renormalization factor $f_\mathrm{ren} = \tau_E/\tau_{E,\mathrm{ISS04}}$, where $\tau_E$ is the energy confinement time and $\tau_{E,\mathrm{ISS04}}$ is the extended International Stellarator Database ISS04 scaling \cite{Yamada2005}.

\begin{figure}
    \centering
    \includegraphics[width=0.45\textwidth]{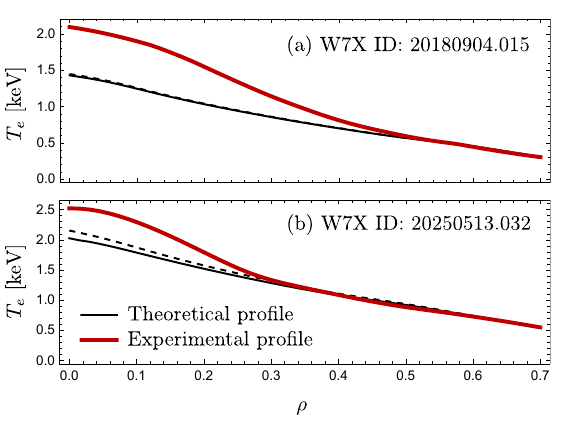}
    \caption{Experimental profiles (fitted) of reference scenarios, compared against theoretical profiles, including (solid) and excluding (dashed) neoclassical contribution to heat flux.  The edge/core boundary conditions for theoretical temperature profiles are set to the experimental values.}
    \label{fig:validation-profiles}
\end{figure}

The first reference discharge corresponds to a gas-fueled electron-cyclotron-heated (ECRH) plasma, with relatively low confinement quality ($f_\mathrm{ren} \sim 0.6-0.7$). The density profile is characteristically flat, and the value of $\eta$ lies in the range of $4$-$5$ across the plasma radius.  Using the computed heat fluxes of W7-X assuming $\eta(\rho) = \eta_\mathrm{exp}(\rho)$ (see Fig.~\ref{fig:heat-fluxes}a), we recover the experimental profile shape of the temperature at outer radial points, as shown in Fig.~\ref{fig:validation-profiles}a, validating our approach to transport modeling; profiles disagree at inner radial points because assumptions of the gyrokinetic simulations are violated experimentally ({\em e.g.} $T_e > T_i$).

The second reference discharge corresponds to a less typical scenario, where improved confinement ($f_\mathrm{ren} \sim 1 - 1.3$) can be achieved in steady state conditions without neutral beams or pellet fueling.  Profiles obtained before peak performance having $\eta \sim 2$ in the neighborhood of $\rho = 0.7$ are used to validate the gyrokinetic transport model for W7-X (Fig.~\ref{fig:heat-fluxes}b), as shown in Fig.~\ref{fig:validation-profiles}b.

{\bf Global confinement estimates.}  Next, we investigate suitable boundary conditions for the SQuIDs using the reference experimental scenarios.  For Stellaris, we first solve the transport equations with edge temperature and density (at $\rho_\mathrm{bc} = 0.7$), and heating $P_\mathrm{ext}$ (localized near $\rho = 0$) set equal to those of the first (high power) reference experimental discharge (Fig.~\ref{fig:validation-profiles}a).  This yields global confinement of $f_\mathrm{ren} \sim 0.50$, with a corresponding profile having $a/L_T \sim 4.6$ at $\rho_\mathrm{bc}$.  Noting that ECRH plasmas are observed to have $a/L_T \sim 4-5$ at this radial location \citep{Ricken-masters-2025}, we propose a second boundary condition, namely $a/L_T = 4$ at $\rho = 0.7$, which leads to a slightly more optimistic prediction of $f_\mathrm{ren} \sim 0.54$ for Stellaris under the same experimental conditions.  These two values, $a/L_T = 4.6$ and $a/L_T = 4$, will be referred to as the conservative and optimistic boundary conditions for Stellaris.

Applying the same procedure for SQuID-$\tau$, except using the low-power/low-$\eta$ reference profiles (Fig.~\ref{fig:validation-profiles}b), yields predictions of $f_\mathrm{ren} \sim 0.70$ and $0.92$ corresponding to conservative and optimistic boundary conditions $a/L_T = 3.5$ and $a/L_T = 3.0$, respectively.  For both SQuIDs, the boundary values of the temperature gradient scale lengths can be used to extrapolate performance to different device scenarios, as observed Gyro-Bohm scaling of experiments implies that the plasma profile shape must be resilient across changes in heating power, and other independent parameters entering empirical scaling laws.

The implications of performance enhancement in SQuID-$\tau$ extend beyond confinement time.  The differences in profile shapes, and in particular the peaking of density that occurs with a particle pinch, compounds with increase in confinement when considering reactor scenarios, as fusion power scales proportional to density squared.  To demonstrate this effect, we solve the transport equations for both configurations including an effective alpha heating source (corresponding to equal parts tritium and deuterium), and applying both the conservative and optimistic boundary conditions.  In all cases the density is set such that its volume average is equal to the Sudo limit \citep{sudo1990scalings}.

\begin{figure}
    \centering
    \includegraphics[width=0.45\textwidth]{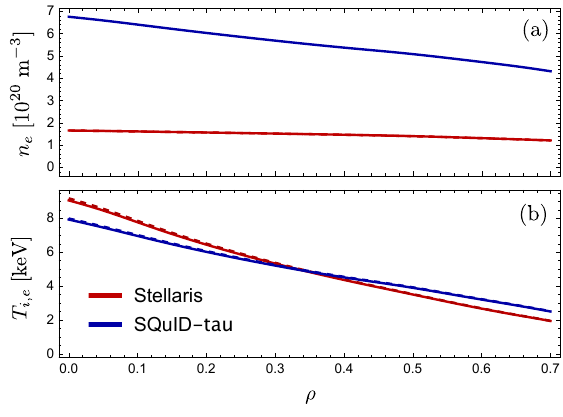}
    \caption{Profiles for high-field ($\langle B\rangle = 10$ T) $Q = 1$ design points, including (solid) and excluding (dashed) the neoclassical contribution to heat transport.  The minor radii of these devices is found to be $a = 0.5$ \commentColor{m} for SQuID-$\tau$ and $a=1.2$ \commentColor{m} for Stellaris.}
    \label{fig:q1-profiles}
\end{figure}

\begin{figure}
    \centering
    \includegraphics[width=0.5\textwidth]{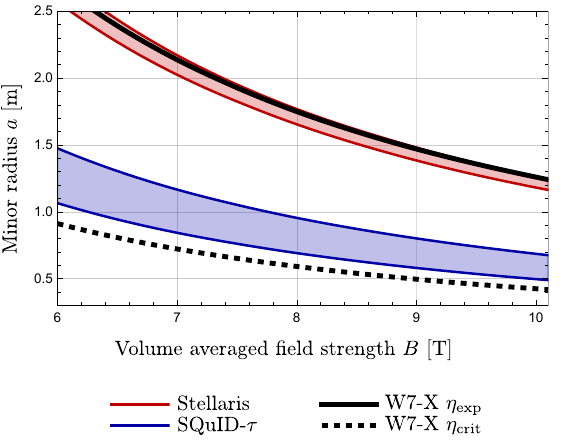}
    \caption{Accessible design points for a stellarator device with fusion gain $Q = 1$.  \commentColor{The colored shaded regions show the range of behavior encompassed by optimistic and conservative boundary conditions.}}
    \label{fig:design-points}
\end{figure}

Varying machine size (minor radius) and field strength (volume-averaged), we obtained estimates for accessible design points for a hypothetical experiment capable of reaching fusion gain $Q = 1$ with total external heating power of $25$ MW deposited near the magnetic axis (see Fig.~\ref{fig:design-points}).  Profiles for a single design point (assuming optimistic boundary conditions) are shown in Figure \ref{fig:q1-profiles}.  For a sample of design points neoclassical heat transport was included but did not contribute significantly (less than $3\%$ of the total heating power for all radial points \commentColor{for SQuID-$\tau$ and Stellaris}); see also Figure \ref{fig:q1-profiles}(b).  Therefore the full diagram was constructed neglecting the neoclassical contribution.  Comparing the results for the two \commentColor{SQuIDs} with optimistic boundary conditions at field strength of $10$ Tesla, a device based on the SQuID-$\tau$ configuration would need to be built with a minor radius of $0.50$ m as compared with a minor radius of $1.18$ m for a device based on the Stellaris configuration.  This corresponds to difference of more than a factor of $13$ in terms of plasma volume, a figure strongly connected to the cost of device construction.  For an alternative perspective, we also consider an ignited reactor scenario achieving $3$ GW of total fusion power with a field strength of $7.5$ Tesla, and an average density of $\sim 63 \%$ of the Sudo limit.  In this case the volume ratio of the two devices exceeds $14$.

\commentColor{For comparison, Figure \ref{fig:design-points} also includes two black curves corresponding to W7-X cases, obtained using transport models based on the data shown in Figure \ref{fig:heat-fluxes}, applying Gyro-Bohm-scaled boundary conditions following the reference experimental profiles (Figure \ref{fig:validation-profiles}).  We must stress that the black curves are only of theoretical interest, since W7-X does not scale to a reactor for a number of reasons, {\em e.g.} insufficient fast particle confinement and large bootstrap current.  The dashed line, in particular, should not really be considered as indicative of possible performance, as the neglect of neoclassical transport is not justified for W7-X.}

Though we focus on electrostatic turbulence in this work, potential limitations due to electromagnetic instabilities such as kinetic ballooning modes (KBMs) must also be considered. To assess this, we performed flux-tube GENE simulations \citep{jenko2000electron} to evaluate KBM stability across the entire core region ($0.2 \leq \rho \leq 0.7$). The analysis was carried out for similar plasma profiles as those cases reported above, but under more ambitious conditions, with local $\beta$ values ranging from $4.2\%$ near the core to $1\%$ at outer radii. Despite the elevated pressure, no KBM instability was observed, suggesting that the underlying magnetic configuration has favorable stability properties, maintaining resilience against KBM drive across the profile, even in high-performance scenarios.

\begin{figure}
    \centering
    \includegraphics[width=1.0\linewidth]{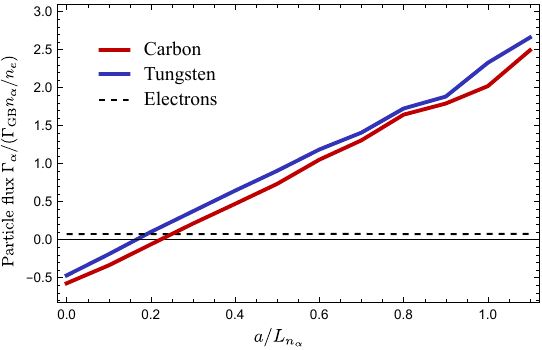}
    \caption{Gyrokinetic particle fluxes of Tungsten ($n_\alpha/n_e = 10^{-3}$, $Z = 44$) and Carbon ($n_\alpha/n_e = 10^{-3}$, $Z = 6$) at $\rho = 0.55$
    , versus normalized density gradient.  Here we assume equal temperature as bulk species, and a fixed background density profile \commentColor{($a/L_n = 0.74$, $a/L_T = 2.0$)}.  Similar result found at other radial locations \commentColor{and gradients}.}
    \label{fig:impurities}
\end{figure}

Another key issue is that of impurity accumulation when turbulence is suppressed by a significant background density gradient \citep{romba2025suppression}.  Encouragingly, gyrokinetic simulations of the transport of both Carbon and Tungsten in SQuID-$\tau$ show a relatively small critical impurity density gradient ($\Gamma_\alpha \approx 0$ at $a/L_{n_\alpha} \sim 0.2$) at all radii.  These assumed a fixed background density profile obtained for a $Q = 1$ scenario, implying that impurities could have a relatively flat profile even when the background gradient is significant; see Figure \ref{fig:impurities}.

In this work we have demonstrated enhanced confinement in the absence of enhanced fueling, using the concept of self-fueling via a strong turbulent particle pinch.  We have therefore restricted our analysis with the conservative assumption of zero turbulent particle fluxes within the plasma volume.  Relaxing this assumption to allow particle sources, we can reasonably expect further gains in confinement -- indeed, we observe strong turbulence suppression in our simulations of SQuID-$\tau$ that have $\eta < \eta_\mathrm{crit}$ (not shown here).  

Predicting global performance in combined ``pinch-pellet'' scenarios, and investigating the full implications for device designs will require detailed further work.  Here SQuID-$\tau$ may turn out to have a significant advantage as compared to W7-X because of its substantially lower neoclassical transport, which is the channel thought responsible for impurity accumulation in W7-X whenever turbulence is sufficiently suppressed \cite{romba2025suppression}.  This issue must however be carefully investigated, both theoretically and experimentally with next-step devices, to find the optimal manner in which to operate a device for truly steady-state and reactor-relevant high performance.

{\bf Acknowledgments.}  We are grateful for discussions with O. Ford and F. Reimold.  We thank S. Bozhenkov, G. Fuchert, J. Knauer and J. Brunner for providing experimental profile data.  We also thank S. Bozhenkov for helpful discussions and feedback on the manuscript.  AGG, PC and PH are supported by a grant from the Simons Foundation (Grant No. 560651)  This work has been carried out within the framework of the EUROfusion Consortium, funded by the European Union via the Euratom Research and Training Programme (Grant Agreement No 101052200 — EUROfusion).  Views and opinions expressed are however those of the author(s) only and do not necessarily reflect those of the European Union or the European Commission.  Neither the European Union nor the European Commission can be held responsible for them.

\bibliography{pinch}

\end{document}